# Oscillator-driven broadband femtosecond transient microspectroscopy with supercontinuum probes generated in a photonic crystal fiber

**RIKUTO FUKUDA,[1] KYOSUKE KISHIHATA,[1] RYO MAN-NAMI,[2] HIROAKI ATSUMI,[2] BORJA CIRERA,[4] DAEGWI KIM,[1,2] AND MASAHIRO SHIBUTA*,[1,2,3]**

[1] *Department of Physics and Electronics, Graduate School of Engineering, Osaka Metropolitan University, 1-1, Gakuen-cho, Naka-ku, Sakai, Osaka 599-8531, Japan*
[2] *Department of Physics and Electronics, Graduate School of Engineering, Osaka City, University, 3-3-138 Sugimoto, Sumiyoshi-ku, Osaka 558-8585, Japan*
[3] *Institute for Molecular Science, National Institutes of Natural Sciences, Okazaki, Aichi 444-8585, Japan*
[4] *Instituto de Ciencia de Materiales de Madrid (ICMM-CSIC), ES-28049 Madrid, Spain*
**Address correspondence to M. Shibuta*
**E-mail: shibuta@omu.ac.jp*

**Abstract:** We present an oscillator-driven femtosecond transient microspectroscopy system based on supercontinuum generation in a photonic crystal fiber (PCF). The system operates at 80 MHz without pulse amplification, enabling high-sensitivity measurements ($<10^{-4}$) while using low pulse energies suitable for microspectroscopy. Optimization of the PCF length (≤50 mm) suppresses dispersion-induced chirp, achieving nearly uniform temporal resolution below 100 fs across 450–900 nm. The high-repetition-rate operation allows rapid spectral acquisition and efficient noise reduction. The developed system provides single-μm scale spatial resolution and supports both reflection and transmission configurations. The ultrafast carrier dynamics in multilayer and monolayer $MoS_2$ are successfully resolved, including spatially dependent relaxation within a single monolayer flake. These results demonstrate a versatile platform for probing local ultrafast dynamics in low-dimensional materials.

## 1. Introduction

Ultrafast optical spectroscopy is an essential tool for investigating photoexcited carrier dynamics in condensed matter and low-dimensional nanoscale materials. In particular, visible-near-infrared (NIR) broadband femtosecond pump–probe spectroscopy enables direct observation of transient electronic and excitonic processes across wide spectral ranges, providing valuable insight into carrier relaxation, energy transfer, and many-body interactions [1–3]. Such measurements have been widely applied to two-dimensional (2D) materials [2,4–6], micro- and nanocrystals [3], organic semiconductors [7], and other cutting-edge functional nanomaterials [8]. However, spatial inhomogeneities, e.g., lattice strain, grain boundaries, and point defects, often give rise to pronounced spatial variations in optical properties and photoexcited carrier dynamics. For example, significant variations in exciton dynamics can occur between the central region, edges, and defect sites even within a single monolayer crystalline flake exhibiting a domain size of 10~100 μm scale [2,6,9–11]. Therefore, spectroscopic techniques capable of resolving ultrafast dynamics with single μm-scale spatial resolution are crucial for understanding local photoexcited processes in targeted materials.

Broadband ultrafast transient spectroscopy is typically implemented using a femtosecond laser combined with a regenerative amplifier that operates at repetition rates of 1–10 kHz [1,12–15]. This is because higher pulse energy more than μJ per pulse is required to generate white-light supercontinuum (SC) probe pulses with nonlinear optical processes in transparent media (e.g., sapphire, YAG, $H_2O$). However, amplifier-based systems have several limitations. For example, high pulse fluence can easily induce sample damage, particularly in tightly focused microspectroscopy configurations. Furthermore, the low repetition rate limits signal averaging efficiency, which becomes problematic when measuring weak transient signals under low excitation conditions.

Oscillator-driven pump–probe spectroscopy is an attractive alternative because it provides high-repetition-rates (~100 MHz) and low pulse energies that are inherently compatible with microspectroscopy and sensitive measurements of fragile materials. However, generating broadband white-light probes directly from oscillator pulses remains challenging because of their low pulse energies.

In this study, we developed an oscillator-driven femtosecond transient microspectroscopy system using SC probe pulses generated in a photonic crystal fiber (PCF). The PCF enables efficient white-light generation even with pulse energies obtained from the oscillator (nJ pulse$^{-1}$), allowing broadband probing without a regenerative amplifier [16–20]. This approach enables ultrafast broadband microspectroscopy with a single μm-scale spatial resolution and high detection sensitivity. The developed system was temporally characterized using a graphite sample to optimize the PCF length. As representative examples, we investigate ultrafast carrier dynamics in multilayer and monolayer $MoS_2$, including spatiotemporally resolved spectroscopic measurements within a single monolayer flake, employing both reflection and transmission (absorption) configurations. These results highlight the versatility of oscillator-driven broadband transient microspectroscopy for studying local ultrafast dynamics in nanoscale materials.

## 2. Experiment

### *2.1. Optical configurations*

Fig. 1 shows a schematic of the high-repetition-rate femtosecond time-resolved microspectroscopy system developed in this study. The system is driven by a femtosecond Ti:sapphire oscillator (Spectra-Physics: Tsunami, 60 fs, 800 nm, 1 W at maximum) with a repetition rate of 80 MHz, where an optical isolator was placed in the beam path to prevent back reflections from the PCF facet into the oscillator, thereby stabilizing the oscillator operation.

The fundamental beam was split into pump and probe lines using a beam splitter (BS) with a 30:70 (pump : probe) ratio. For the pump line, the beam was focused by a concave mirror ($f$ = 50 mm) into a β-barium borate (BBO) crystal (thickness: $t$ = 1 mm) to generate the second harmonic (400 nm) for the pump pulse. The fundamental 800 nm pulse can also be used as the pump pulse by removing the BBO crystal. In the probe line, a PCF (NKT Photonics: NL-PM-750, see inset for cross section of the PCF [21]) was used to generate broad band SC pulses, in which the fundamental beam was coupled using an aspherical lens (Thorlabs: A375TM-B, $f$ = 7.5 mm, NA = 0.30). This results in efficient SC generation with a high-order nonlinear optical process without pulse amplification [22,23]. The generated SC light covered a targeted visible-NIR range of 450–900 nm and served as the probe pulse (Fig. 2). The output SC beam was collimated using an objective lens (Newport : 40×).

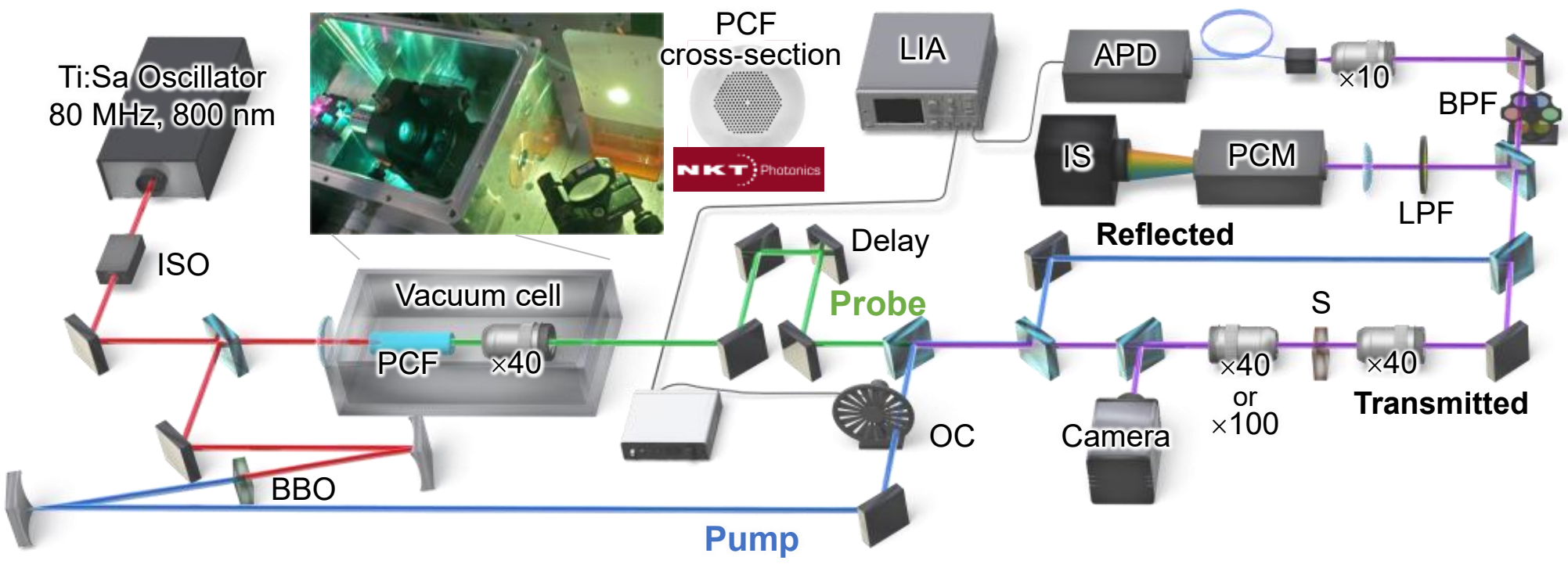

**Fig. 1.** Schematic of the high-repetition-rate femtosecond transient microspectroscopy system. Femtosecond Ti:sapphire oscillator (800 nm, 100 fs, 80 MHz) was split into pump and probe lines. The pump line was focused onto a β-BBO crystal to generate a 400 nm pump pulse. The probe line was coupled to a PCF (inset: cross section of PCF [21]) enclosed in a vacuum cell (see picture) to stably generate broadband SC probe pulses. Both pump and probe pulses were recombined and focused onto the sample using an objective lens (×40 or ×100). The transmitted or reflected probe pulse was then filtered by a long-pass filter (LPF) to remove the pump pulses. The spectral profile of the probe pulse is detected using a polychromator (PCM) and a linear imaging sensor (IS), and the modulation of the probe spectrum with and without pump pulse is obtained by synchronizing an optical chopper (OC, see the timing chart in Fig. 2). Single-channel detection using an avalanche photodiode (APD) and a lock-in amplifier (LIA) is also available for a selected wavelength. Other abbreviated components; BS: beam splitter, BPF: band-pass filter, DM: dichroic mirror, and S: Sample.

The length of the PCF ($L_{\mathrm{PCF}}$) was set to 30–50 mm, which was optimized based on a temporal characterization described later (see Section. 3. 1.). Compared with conventional SC generation using solid or liquid materials, which require much more than 1 μJ/pulse, the PCF allows efficient broadband SC generation with a mode-locked laser oscillator with significantly lower pulse energy. In this system, an input pulse energy of ~8 nJ/pulse produced a sufficiently broad SC output.

The fiber input facet can cause thermal damage in ambient air due to the tight focusing of the fundamental pulse into the PCF, where such degradation results in narrowing and less stability of the SC spectral profile. To solve this issue, the PCF and collimation objective lens (×40) was enclosed in a vacuum cell (~1 Torr), which effectively suppressed the damage to the fiber facet. Therefore, the vacuum-sealed configuration enables the long-term stability of the SC probe pulses for more than 100 h without noticeable degradation. Although end-cap processing is often used to improve the durability of PCFs, it is not always practical for the short PCF ($L_{\mathrm{PCF}}$ = 30–50 mm) required in the present system. Because shortening the PCF is essential for suppressing dispersion-induced temporal broadening, we employed the vacuum cell as a simple and effective way to protect the fiber facet while preserving the optimized short-PCF configuration.

Note that the PCF used in this study is designed for the Ti:sapphire oscillator. However, the present approach can be readily extended to other femtosecond oscillator systems by selecting PCFs tailored for different wavelength ranges, such as those compatible with Yb-based femtosecond lasers [18].

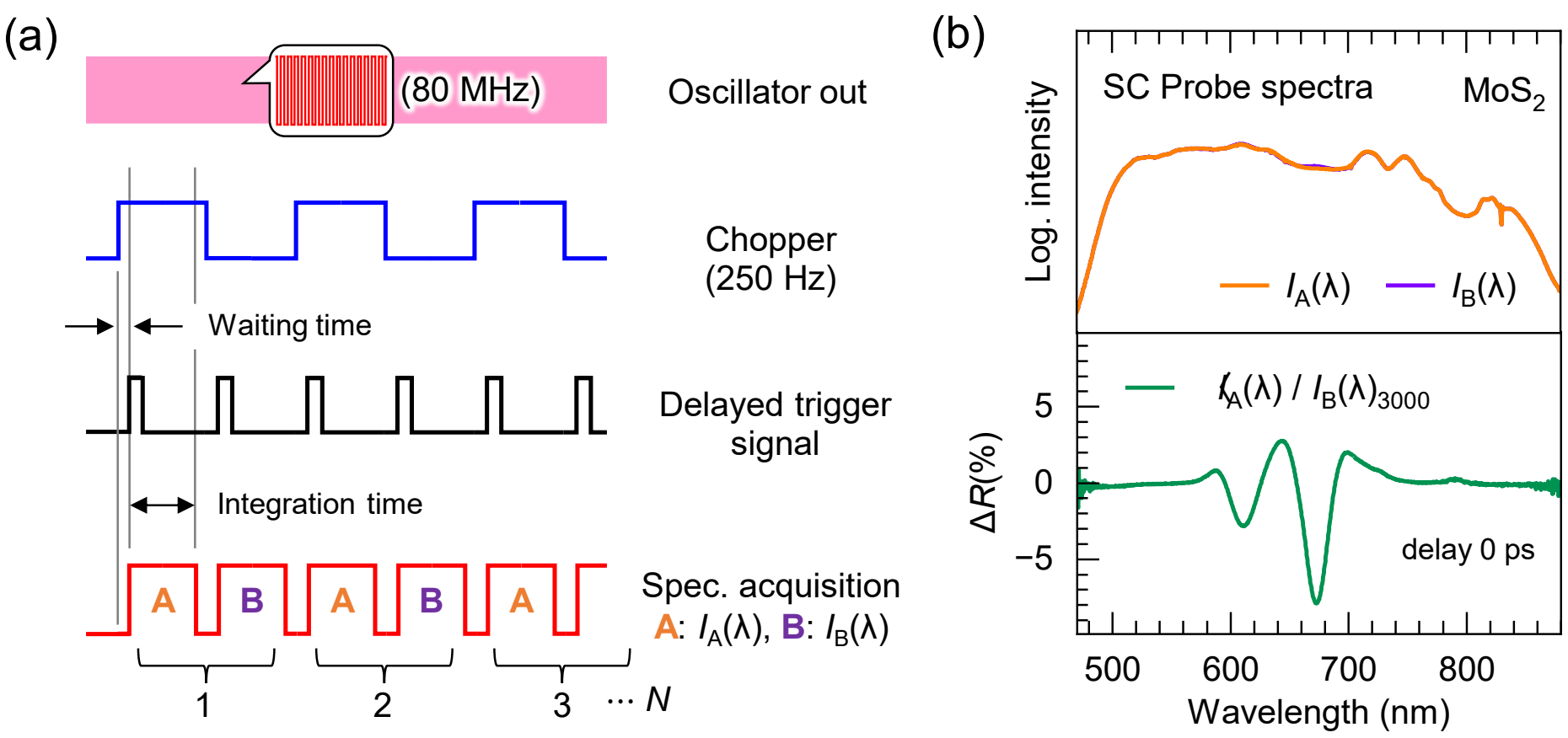

**Fig. 2.** (a) Timing chart for spectral acquisition. The intensity of the pump beam was modulated at 250 Hz using an optical chopper. The rising and falling edges of the chopper's synchronization signal with an appropriate waiting time (typ. 100 μs) are used to trigger the data acquisition system. Probe spectra corresponding to the pump-on and pump-off conditions are alternately accumulated as A: $I_A(\lambda)$ and B: $I_B(\lambda)$, respectively. These spectra are mostly overlapped because the pump-induced spectral change is small. The accumulation time was set to 1.8 ms in typical. The differential acquisition and integration over $N$ times suppress spectral fluctuations of the SC probe. (b) Probe spectra with pump excitation $I_A(\lambda)$ and without excitation $I_B(\lambda)$. (c) Averaged spectra of the intensity ratio $\langle I_A(\lambda)/ I_B(\lambda)\rangle_N$ obtained from accumulation over $N$ cycles.

The temporal delay between the pump and probe pulses was controlled using a mechanical delay stage (SIGMAKOKI FS-3150X) with a minimum step of 100 nm (0.33 fs). After adjusting the light fluence using neutral density filters, both pump and probe beams were recombined collinearly using a dichroic mirror (DM) and focused onto the sample through a microscope objective (×40 or ×100). The spot size governs the spatial resolution of the system, which reaches a single-μm scale (see Fig. 6(b)). An optical microscope system was integrated into the setup to monitor the pump and probe beam on the sample. An LED light was used to image the sample at whole field of view (not shown in Fig. 1).

The optical layout allows both transmission and reflection measurements, enabling time-resolved transient spectroscopy of transparent and opaque samples. After removing the pump pulse using a long-pass filter (LPF; in case of a 400 nm pump), the spectral profile of the probe pulse was measured using a polychromator (SOL instruments SL100M, $f$ = 100 mm, $F$/3.3) and a linear imaging sensor (Hamamatsu S11639-01) driven by an electric circuit (Hamamatsu C16605).

### *2.2. Data acquisition.*

A synchronized detection system was constructed to acquire the transient spectral profile under pump pulse illumination. The timing chart of the detection system is shown in Fig. 2. The pump beam is modulated at 250 Hz (4 ms cycle, duty 50%) by an optical chopper where the reference signal was synchronized into a digital delay generator (Stanford Research Systems: DG535) generating trigger signal with an appropriate waiting time (100 μs in typ.) for spectral acquisition. Then, the probe spectra with $I_A(\lambda)$ and without pump excitation $I_B(\lambda)$ were alternately recorded every 2 ms, with an accumulation time of typically 1.8 ms. The intensity ratio, $I_A(\lambda)/I_B(\lambda)$ was accumulated $N$ cycles to obtain the averaged spectra $\langle I_A(\lambda)/ I_B(\lambda)\rangle_N$. The signal-to-noise (S/N) ratio depending on the $N$ value is discussed in Section 3. 2. This accumulation system allows us to obtain weak transient optical responses down to $<10^{-4}$ even using PCF-based probe pulses which are generally more fluctuating than those obtained by solid or liquid media at an amplifier-based light source.

We also use a single-channel lock in detection to check the spatiotemporal overlap between the pump and probe pulses as an initial optimization (top-right in Fig. 1). In this case, an avalanche photodiode (APD: Hamamatsu C12702-11) combined with a lock-in amplifier (Stanford Research Systems SR810) was used to detect the transient optical signal of the probe pulse. The reference modulation for lock-in detection was produced by another optical chopper capable of operation at 70 kHz for single channel-detection. This detection system can also be used to obtain an intensity trace at a specific wavelength by an insertion of bandpass filter (BPF).

## 3. Results and Discussion

### *3.1. Temporal characterization of PCF-based transient microspectroscopy*

As mentioned in the experimental section (Section 2), the PCF is indispensable for generating SC to be used as a broadband probe pulse in the oscillator-driven transient microspectroscopy. In the SC generation process, the fundamental and generated probe pulses propagate through the $SiO_2$ fiber core [21] over a length much longer than that typically used in bulk solid or liquid media in amplifier-based systems. Therefore, the temporal characterization and optimization of the developed system are essential to obtain the best performance for the transient spectroscopy in terms of time resolution. The length of the PCF, $L_{PCF}$, is one of critical parameters because the group velocity dispersion (GVD) must affect the temporal distribution of the SC probe pulse in each spectral region.

To evaluate the pulse characteristics across the entire SC spectral range, a highly oriented pyrolytic graphite (HOPG) was used as a reference material. HOPG exhibits a visible ultrafast optical response that occurs nearly instantaneously after photoexcitation [24–26], making it suitable for characterization of the PCF-based transient microspectroscopy in time domain. Fig. 3(a) shows the pump-probe intensity traces of the transient reflection ($\Delta R$) measured by the single-channel detection using BPFs (550–800 nm with 50 nm intervals, transmission band; 10 nm) with $L_{PCF}$ = 130 mm. Clear transient optical responses were observed at all probe wavelengths, where they systematically shifted in the relative pump-probe delay. The tendency clearly indicates that the longer wavelength component reaches the sample earlier than the shorter wavelength component. This shift is attributable to the group delay introduced during propagation of the SC pulses in the PCF.

Next, we examined the dispersion-induced shift at different $L_{PCF}$. Fig. 3(b) summarizes the shift of time origin for $L_{PCF}$ = 35, 50, 130, and 300 mm (the peak at each wavelength) with respect to that at a probe wavelength of 550 nm. The amount of shift between 550 and 800 nm decreases as $L_{PCF}$ becomes shorter. However, even for the shortest $L_{PCF}$ ($L_{PCF}$ = 35 mm), the total shift in the wavelength range remains almost the same (4.3 ps) as that obtained at $L_{PCF}$ = 50 mm.

Furthermore, we evaluated the system response function (RF) for each spectral range extracted from the deconvolution of the intrinsic decay component of HOPG (time constant = 200 fs) [24]. The full width at half maximum (FWHM) of the RF was plotted against the probe wavelength, as shown in Fig. 3(c). The RF width value is significantly narrower when a shorter $L_{PCF}$ is employed, whose effect is especially significant at a shorter wavelength. This behavior is reasonably explained by the PCF design, provided that the zero-GVD wavelength is 750 nm [21]. Therefore, pulse broadening (i.e., chirping) due to GVD becomes significant at longer $L_{PCF}$ and shorter wavelengths further away from 750 nm. Notably, the RF width is not significantly affected by further shortening to less than 50 mm.

From these results, we optimized the $L_{PCF}$ in the present system to be $L_{PCF} \leq 50$ mm, which can minimize both the group delay and temporal broadening in the whole spectral range of the SC probe pulse while maintaining sufficient conversion efficiency of the SC generation. At $L_{PCF} \leq 50$ mm, the time resolution at the whole spectral range is estimated by < 100 fs; a deconvolution analysis can be used to extract the time constant shorter than the RF width. Note that because the pulse duration of pump beam (400 nm) is considerably shorter (ca. 100 fs) than the SC probe pulse, the contribution of the pump pulse to the present temporal resolution was negligible.

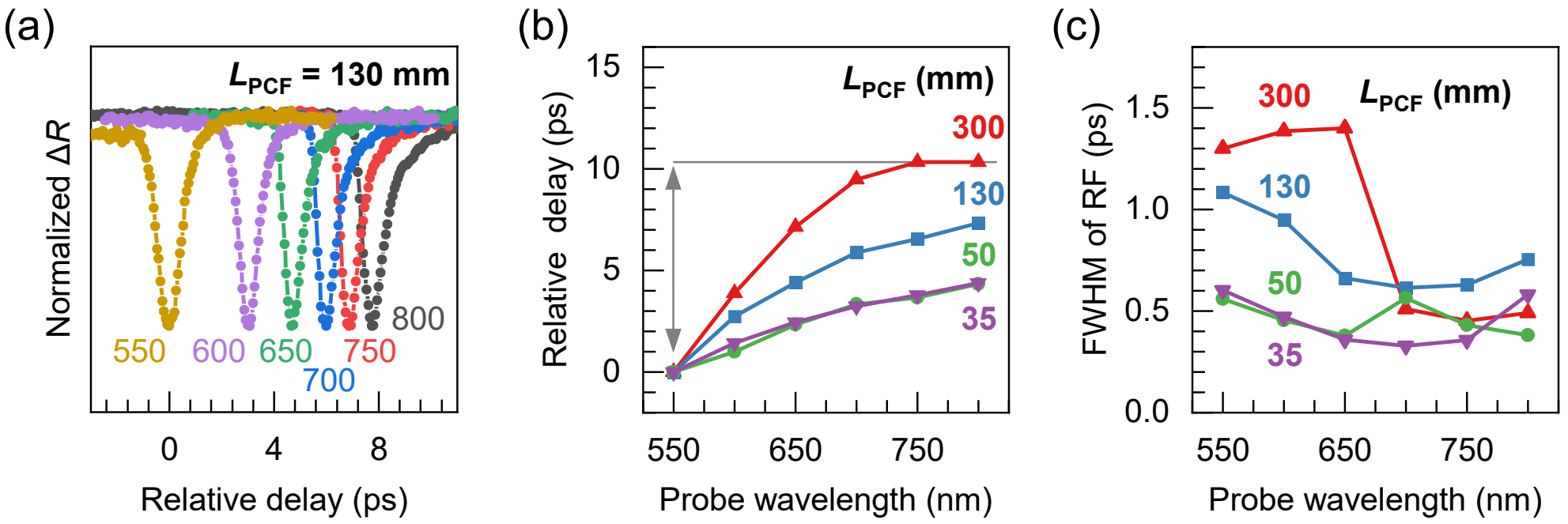

**Fig. 3.** Temporal characterization of the SC probe generated by the PCF. (a) Transient reflectance traces of HOPG ($L_{PCF}$ = 130 mm) against relative pump-probe delay at selected probe wavelengths. (b) Delays in the intensity maximum at each probe wavelength for different $L_{PCF}$, showing that shortening the $L_{PCF}$ effectively suppresses the amount of shift between 550 and 800 nm (vertical arrow). (c) Wavelength dependence of the system RF in FWHM.

### *3.2. Noise reduction by averaging the transient spectra.*

Since the PCF-based SC generation generally exhibits higher spectral fluctuation than that generated in solid or liquid media using amplifier-based light sources, the S/N ratio versus accumulation time should be evaluated. Fig. 4(a) shows the $\langle I_A(\lambda)/ I_B(\lambda)\rangle_N$ for various $N$ values obtained *without* pump pulse. Namely, we will obtain $\Delta R = 0\%$ at whole spectral range in an ideal averaging. The actual acquisition time can be calculated by $N \times 4$ ms/cycle (Fig. 2).

As the $N$ value increases, the apparent intensity fluctuations of the SC probe spectra are progressively suppressed. Previous studies using regenerative amplifier systems have reported that typical transient optical responses in 2D materials exhibit signal changes on the order of ~0.1% at typical excitation fluences, $F$, around 30 μJ cm$^{-2}$ [27]. In this system, the $\langle I_A(\lambda)/ I_B(\lambda)\rangle_{1000}$ (4 s acquisition) provides sufficient sensitivity to reliably detect such signals. The present quick spectral averaging owing to high-repetition-rate operation indicates that our oscillator-driven optical system is capable of transient spectroscopy, even using PCF-based SC light.

The S/N ratio is further improved by increasing the $N$ to 10000 (40 s acquisition), enabling highly sensitive measurements. The inset in Figure 4 plots the root-mean-square (RMS) noise evaluated in the wavelength range between 500 and 800 nm against $N$, where it was calculated as the standard deviation of the $\langle I_A(\lambda)/ I_B(\lambda)\rangle_N$. The RMS noise decreases proportionally to $N^{-1/2}$, which is in excellent agreement with the expected statistical behavior for uncorrelated noise. Assuming that the detection limit is defined as $3\sigma$ [28], the proposed system achieves a detection sensitivity of approximately 0.021% for $N$ = 10000. When setting $N$ = 30000, the detection sensitivity is extrapolated to be <$10^{-4}$ (0.009%). This result demonstrates that high-repetition-rate averaging effectively suppresses the fluctuations in the probe intensity and enables the detection of small transient optical responses.

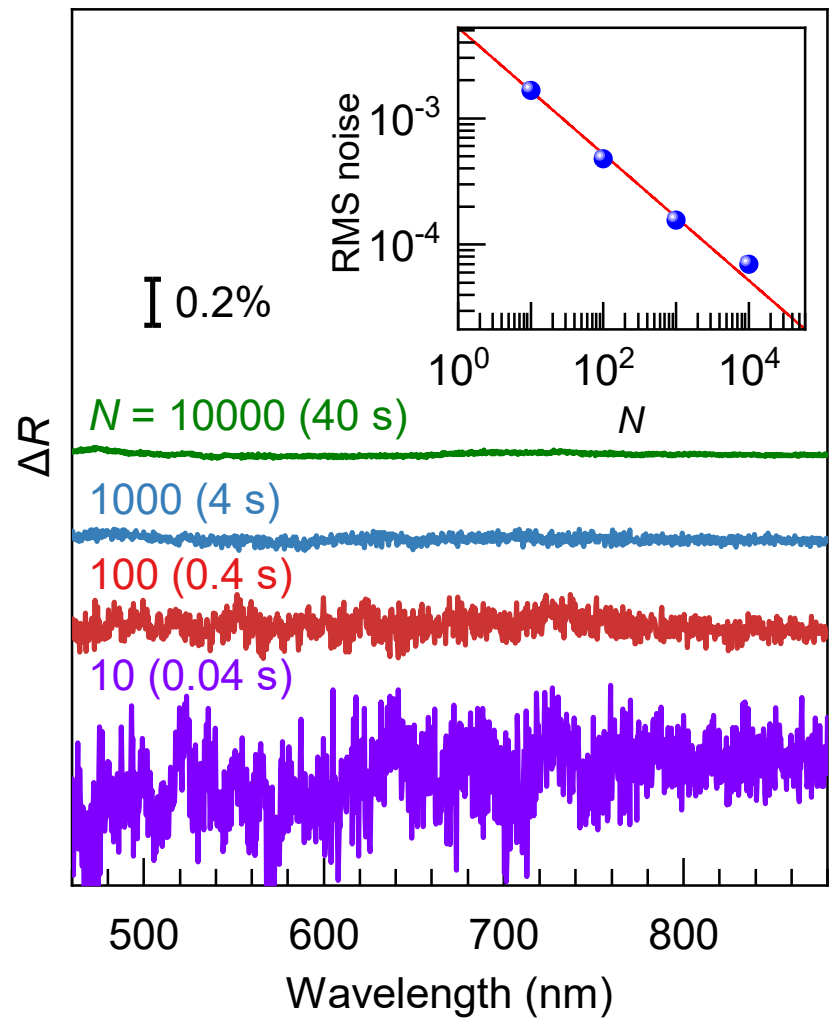

**Fig. 4.** Suppression of probe spectrum fluctuations. (a) Averaged $\Delta R$ spectra ($\langle I_A(\lambda)/I_B(\lambda)\rangle_N$) for the HOPG without the pump pulse. With increasing $N$ (acquisition time). The spectral fluctuation is effectively suppressed. (inset) RMS noise in the spectral region between 500 and 800 nm. The RMS decreases approximately as $N^{-1/2}$, indicating that their statistical fluctuations are dominated by the measurement noise and can be efficiently reduced.

### *3.3. Chirp correction*

As shown in Fig. 3, the group delay induced by light traveling in the PCF is unavoidable even with the shortest $L_{PCF}$, which is a common technical issue with amplifier-based systems. In this section, we describe the chirp correction procedure for PCF-generated SC light using the HOPG sample. Fig. 5(a) shows the 2D-transient reflectance spectrum of the HOPG. A systematic shift in the apparent rising edge of the optical response signal is observed depending on the probe wavelength due to the group delay of the SC light (Fig. 3(a,b)).

To remove the effect, the wavelength-dependent shift of the time origin, $\Delta t_0(\lambda)$, was determined from the raw data (Fig. 5(a)). A time compensation was then performed so that the $\Delta t_0(\lambda)$ appears at the zero-time delay. Fig. 5(b) shows the corrected transient reflectance spectrum for the data shown in Fig. 5(a), confirming that the dispersion-induced chirp of the SC probe has been successfully compensated. Although $\Delta t_0(\lambda)$ function sensitively depends on the $L_{PCF}$ used, the corresponding $\Delta t_0(\lambda)$ function can be readily re-determined through the same calibration measurement that takes approximately 15 min. All the data presented below have been corrected using this calibration method.

### *3.4. Transient reflection microspectroscopy for multilayer $MoS_2$*

To demonstrate the capability of the present oscillator-driven optical system, we performed femtosecond transient microspectroscopy on $MoS_2$, a representative two-dimensional transition-metal dichalcogenide (TMD). TMDs have attracted considerable attention because of their strong light–matter interaction and excitonic/photonic effects [29–33]. A multilayer $MoS_2$ sample was prepared by mechanical exfoliation. Fig. 6(a) shows the atomic force microscopy (AFM) image of prepared $MoS_2$ flake, where the thickness is estimated to be 20-50 nm, roughly 30-80 $MoS_2$ layers. The photoexcited carrier dynamics were investigated under the reflection configuration because of its low transmittance.

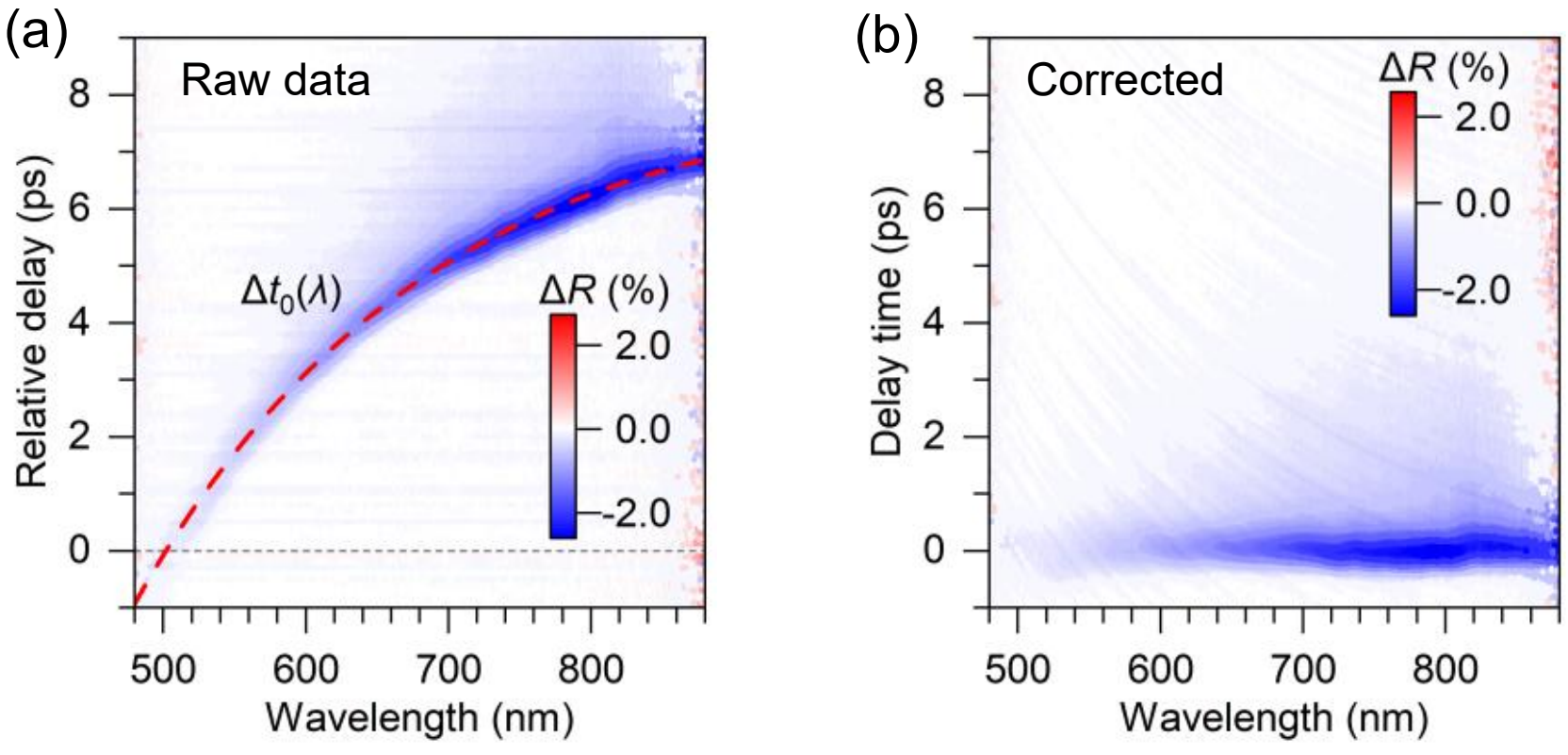

**Fig. 5.** Correction of the wavelength-dependent difference in the time origin. (a) Measured 2D transient reflectance map of the HOPG before and (b) after correction.

The pump and probe pulses were focused on the central region of the flake (marked in Fig. 6). The intensity line profile of both pulses (Fig. 6(b)) indicates that both beams are tightly focused onto the sample about a single μm in diameter depending on the objective lenses. Using ×40 (or×100) objective lens, the spot diameters (FWHM) of the pump and probe pulses were 1.9 μm (or 1.0 μm) and 2.3 μm (or 0.6 μm), respectively. Since the detected transient signal originates only from the region excited by the pump beam, the spatial resolution of the present microspectroscopy corresponds to the pump pulse diameter, i.e., 1.95 μm for ×40 and 0.51 μm for ×100, applying a 16-84 intensity criterion.

Fig. 6(c) shows the broadband transient reflectance ($\Delta R$) spectrum obtained for the multilayer $MoS_2$ ($F$ = 260 μJ cm$^{-2}$). Even with a short time accumulation ($N$ = 3000, 12 s), the spectral dataset was obtained with a sufficient S/N ratio. The measured transient spectrum shows pronounced negative signals at approximately 670 and 610 nm, which originate from the ground-state bleaching (GSB) of excitons known as the A and B excitons (A- and B-GSB) in $MoS_2$, respectively. These excitonic resonances correspond to direct optical transitions; the energy separation between the A and B excitons arises from spin–orbit splitting of the valence band [34–36].

In addition to the GSB signals, positive $\Delta R$ signals appear on both sides of the negative GSB signal. These features are mainly attributed to an excited-state absorption (ESA) arising from photoexcited carriers (see also Fig. 7(d) for energy diagram), broadly distributed in the whole spectral range from 580 to 760 nm. Specifically, optical excitation with a 400 nm pump pulse induces an initial transition to high-energy states, such as the C exciton originating from band nesting, which subsequently gives rise to the observed ESA (as C-ESA) [37,38]. These spectral features obtained in our oscillator-driven system are consistent with those previously reported for multilayer $MoS_2$ using amplifier-based laser systems [27,39].

For the analysis, $\Delta R$ intensities at specific wavelengths (integrated by 5 nm in width) for all transient spectra (180 in total: 150 steps of 0.2 ps/step for the delay from −6 to 24 ps, and 30 steps of 1 ps/step for the delay after 24 ps) are plotted against the delay time, as shown in Fig. 6(d), further demonstrating the stability of the present system during the transient microspectroscopy. These traces were fitted using single- or double-exponential functions (as lifetimes: $\tau_1, \tau_2$ and their amplitudes: $A_1$, $A_2$), as summarized in Table 1. These excited-state dynamics persisting in the order of sub-ns are consistent with reported ones for $MoS_2$ using amplified laser systems [27,39,40].

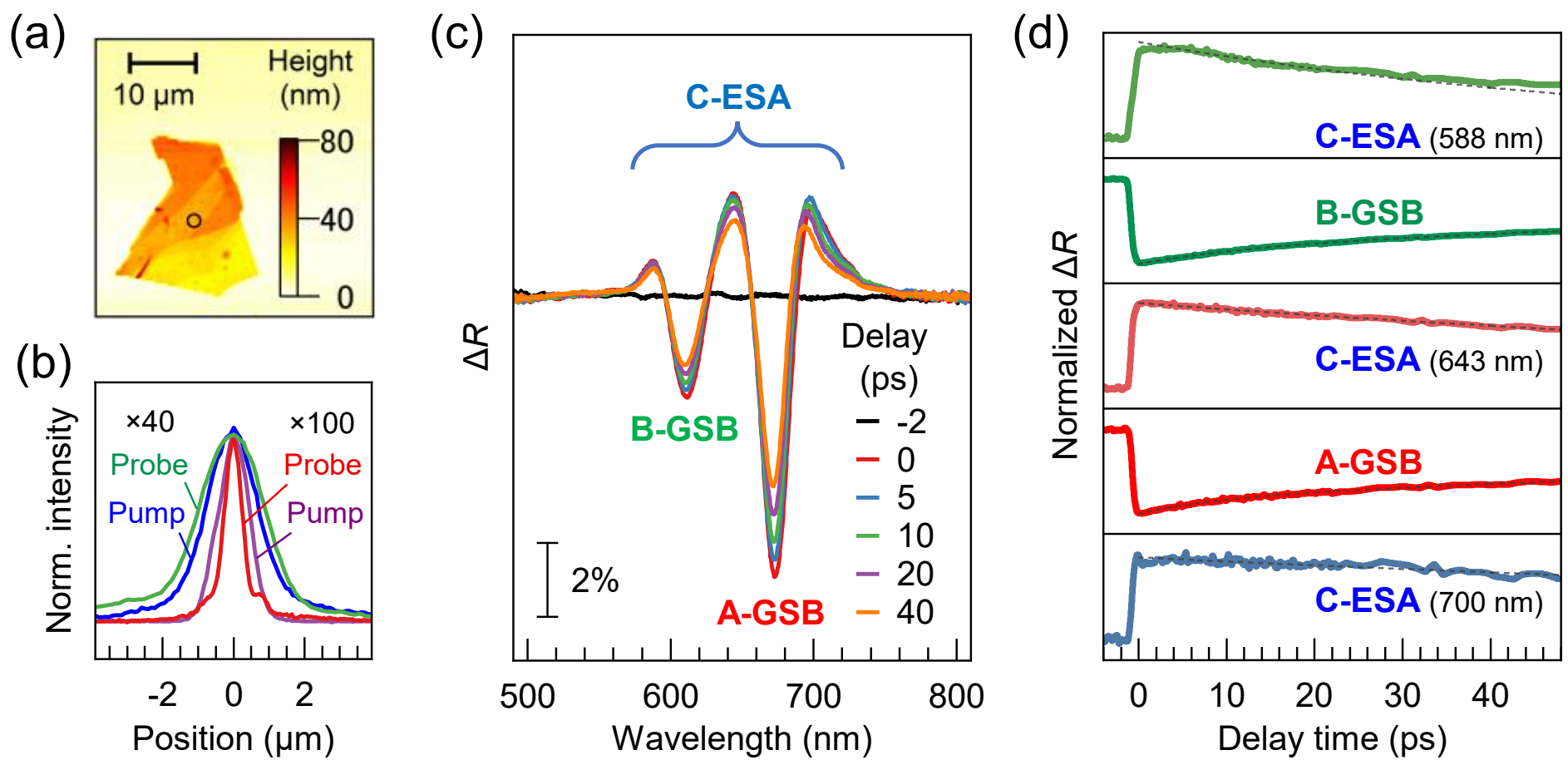

**Fig. 6.** Transient reflectance microspectroscopy of a multilayer $MoS_2$. (a) Atomic force microscopy image of the prepared $MoS_2$. The thickness was approximately 20-50 nm. (b) Intensity profiles of the pump and probe beams. Using a ×40 (or ×100) objective lens, the spot sizes are 1.9 μm (or 1.0 μm) for the pump and 2.3 μm (or 0.6 μm) for the probe beams. The spatial resolutions determined from the probe profiles are 1.95 μm for ×40 and 0.51 μm for ×100 objectives, based on a 16-84 intensity criterion. (c) Transient reflectance spectrum for $MoS_2$. The negative signals at approximately 670 and 610 nm originate from grand state bleaching of A- and B-excitons (A- and B-GSB). Positive signals are attributed to the excited-state absorption of C-exciton (C-ESA), which is distributed over a spectral range of 580-760 nm. (d) Intensity traces at representative probe wavelengths indicated by the colored triangle in (c) (integrated 5 nm in width). The dashed lines represent the results of the exponential fitting using single- or double-exponential functions. The fitting parameters are summarized in Table 1.

**Table 1. Fitting parameters of transient microspectroscopy for multilayer $MoS_2$.**

| Wavelength (nm) | Assignment | $A_1$ | $\tau_1$ (ps) | $A_2$ | $\tau_2$ (ps) |
|---|---|---|---|---|---|
| **588** | C-ESA | 0.98 | 198.1 | - | – |
| **612** | B-GSB | 0.82 | 167.4 | 0.18 | 9.61 |
| **643** | C-ESA | 1.01 | 127.8 | – | – |
| **673** | A-GSB | 0.80 | 183.8 | 0.21 | 14.7 |
| **700** | C-ESA | 1.14 | 61.1 | – | – |

### *3.5. Transient absorption microspectroscopy for monolayer $MoS_2$*

To demonstrate that the developed system can also be applied to atomically thin films in a transmission configuration, we employed monolayer $MoS_2$ synthesized by chemical vapor deposition (CVD) on a sapphire substrate. Fig. 7(a) shows the local transient absorption ($\Delta A$) spectrum measured for the monolayer $MoS_2$. The $F$ value of 104 μJ $cm^{-2}$ was kept comparable to values commonly used in previous studies using amplifier-based systems without tight focusing.

The transient optical signal of the monolayer $MoS_2$ is much smaller ($\Delta A$ <0.5%) than that of the multilayer sample at the reflection configuration (Fig. 6). Nevertheless, we obtained a high-quality dataset by taking the accumulation time of 40 s per spectrum ($N = 10000$). The 2D spectral map generated from 85 spectra (75 steps of 0.5 ps; delay from −4.5 ps to 33 ps, and 10 steps of 2 ps; delay after 33 ps delay) shown in Fig. 7(b) clearly demonstrates that high sensitivity and stability can be maintained while acquiring broadband transient spectra for one to several hours.

The transient absorption spectra are similar to the transient reflection spectra obtained for the multilayer $MoS_2$ (Fig. 6). This similarity arises because pump-induced changes in the absorption coefficient near the excitonic resonances dominate the transient reflectance response. The measured spectra exhibit negative signals with A- and B-GSB and positive signals with C-ESA, as investigated in the multilayer $MoS_2$. The peak positions of GSB agree well with those observed in the steady-state absorption spectrum (upper panel in Fig. 7(a)). Unlike the multilayer $MoS_2$, the relaxation of the excited state occurs at a shorter time scale. Notably, the spectral feature due to C-ESA at the longer wavelength side (≥ 680 nm) shifts toward the shorter wavelength.

Figure 7(c) summarizes the apparent peak wavelength and integrated intensity of the C-ESA at the wavelength of 678-770 nm as a function of delay time. It is known that the observed blue shift is attributed to the cooling of the hot carriers toward the band-edge states related to the C exciton [41]. Namely, just after photoexcitation, the transient signal originates from the ESA of hot carriers into higher-lying bands with larger effective masses. As the carriers dissipate their excess energy through carrier–phonon scattering, the ESA spectral feature progressively shifts toward shorter wavelengths, as illustrated in Fig. 7(d). The rate of this spectral shift slows and nearly saturates at a delay time of 20 ps, indicating that the hot-carrier cooling process is mostly completed within sub-100 ps timescale.

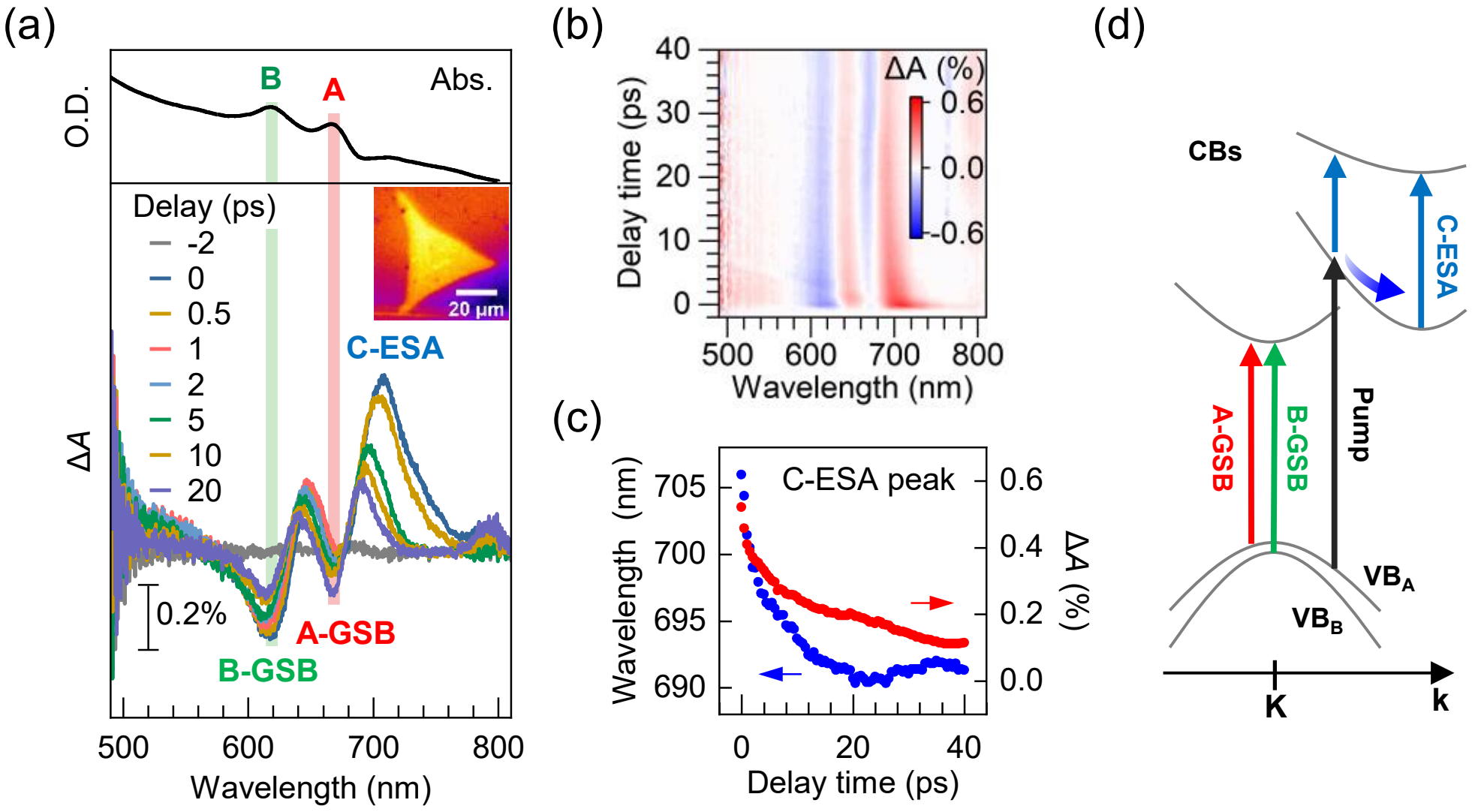

**Fig. 7.** Ultrafast photoexcited state dynamics of monolayer $MoS_2$. (a) Transient absorption spectra (pump 400 nm, $F$ = 104 μJ cm$^{-2}$), showing the A-GSB, B-GSB, and C-ESA signals. (upper panel) The steady-state absorption spectrum of monolayer $MoS_2$. The ESA feature at ≥ 680 nm blue-shifted as the delay time increases. (b) 2D spectral map as a function of the probe wavelength and delay time, demonstrating broadband measurements with high S/N ratio. (c) Delay-time dependence of the peak positions and integrated intensity of the C-exciton ESA signal at 678−770 nm. The observed blue shift indicates ultrafast energy relaxation from hot carriers toward the C-exciton band edge within several tens of picoseconds. (d) Schematic of the energy diagram and carrier dynamics relevant to the A, B, and C excitons [41].

Previous studies have discussed the hot-carrier relaxation in C excitons using amplifier-based pump–probe spectroscopy [41–43]. However, it remains challenging to directly track the ultrafast spectral evolution of the C-ESA with a high S/N ratio over a broad spectral range. The present measurements clearly resolve this dynamic spectral evolution across the entire probe bandwidth, highlighting the capability of the developed high repetition-rate transient microspectroscopy system to sensitively capture subtle ultrafast carrier dynamics in 2D materials. This capability is particularly important for atomically thin materials, where ultrafast carrier relaxation processes are often spatially heterogeneous, as discussed in the next section.

### *3.6. Spatially resolved ultrafast dynamics in a single $MoS_2$ monolayer flake*

Finally, we investigated the spatial dependence of the photoexcited carrier dynamics within a single $MoS_2$ monolayer. In CVD-grown monolayer $MoS_2$ crystals, lattice strain and point defects accumulate preferentially near the edges of the flake. Previous studies have reported that local optical properties—such as photoluminescence and Raman spectra—can differ significantly between the center and edge regions of the flakes [9–11].

However, these optical measurements cannot directly reveal the spatial variations in ultrafast carrier dynamics. Furthermore, many studies probing local femtosecond dynamics have so far relied on single-channel detection at a fixed probe wavelength. In contrast, the system developed in this work enables spatially resolved measurements of broadband transient absorption spectra, allowing the investigation of ultrafast carrier dynamics across the entire spectral range with a spatial resolution better than 1 μm.

Fig. 8 shows the transient absorption spectra measured at the (a) center and (b) near-edge regions of a CVD-grown single $MoS_2$ flake. The characteristic spectral features associated with the A-, B-GSB, and C-ESA signals are clearly observed in both regions, indicating that the overall excitonic structure is preserved across the flake.

Despite this similarity in spectral features, the temporal evolution of these signals exhibits clear position dependence. At the center of the flake, the transient signals decay relatively slowly, showing a dominant relaxation component on the order of several tens of picoseconds. In contrast, near the flake edge, the signals decay more rapidly and approach the baseline within a shorter time scale. These results indicate that even within a single crystalline flake, the photoexcited carrier dynamics vary depending on the probing position.

Fig. 8(c) shows the position-dependent time traces at the characteristic wavelengths due to A-, B-GSB, and C-ESA signals. The relaxation dynamics were analyzed by fitting the traces with single- or double-exponential functions, and Table 2 summarizes the resulting fitting parameters. Quantitative analysis revealed pronounced differences in the relaxation dynamics between the center and edge regions for the A and C excitons. The long-lived components ($\tau_2$) of the A- and C-exciton dynamics in the central region were found to be 75.2 and 45.6 ps, respectively, whereas they were largely suppressed at the flake edge to have only short-lived ($\tau_1$) component. This behavior is likely related to enhanced recombination or carrier-trapping channels at structural defects, which are expected to be more abundant near the flake boundaries.

The difference between the center and edge regions is less pronounced for the B exciton. Although the relaxation dynamics exhibit a considerable acceleration, long-lived component ($\tau_2$) remains near the edge, whose contrasted result may be related to differences in the spatial extent and diffusion properties of the various excitonic states. The ability to acquire broadband transient spectra at specific positions within a single $MoS_2$ flake highlights the unique capability of this system for investigating local photoexcited processes in low-dimensional materials.

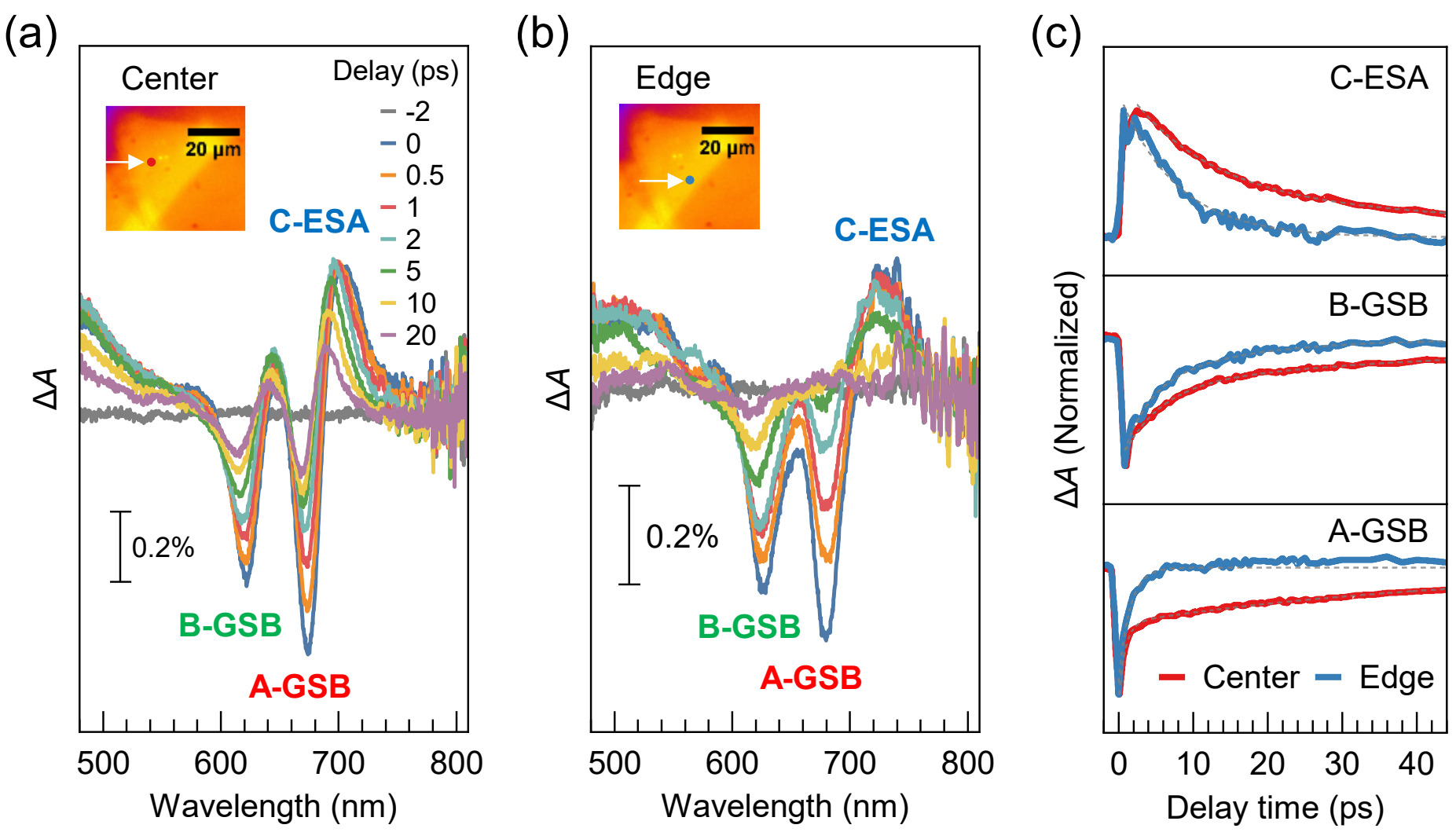

**Fig. 8.** Spatially resolved transient absorption dynamics of a single $MoS_2$ monolayer. (a,b) Transient absorption spectra measured at the (a) center and (b) near edge. The insets in panels (a) and (b) indicate the measured positions. (c) Time-resolved intensity traces extracted at the peak wavelengths corresponding to the (bottom) A-GSB, (middle) B-GSB, and (top) C-ESA signals, respectively.

**Table 2. The fitting parameters obtained from the exponential analysis of the relaxation dynamics revealed significant differences between the center and edge regions.**

| Signal | Position | $A_1$ | $\tau_1$ (ps) | $A_2$ | $\tau_2$ (ps) |
|---|---|---|---|---|---|
| **A-GSB** | center | 0.56 | 0.85 | 0.44 | 75.2 |
| | edge | 0.90 | 1.4 | | |
| **B-GSB** | center | 0.51 | 5.3 | 0.36 | 55.4 |
| | edge | 0.48 | 3.5 | 0.40 | 13.8 |
| **C-ESA** | center | 0.74 | 11.3 | 0.31 | 45.6 |
| | edge | 1.04 | 7.8 | | |

## 4. Conclusion

In summary, we have developed a high-repetition-rate femtosecond transient absorption and reflectance microspectroscopy system driven by a Ti:sapphire oscillator. By using a PCF for a broadband SC generation, the system realizes amplifier-free pump–probe spectroscopy over the 450–900 nm spectral range. By optimizing the $L_{PCF}$ to be ≤ 50 mm based on the temporal characterization, uniform temporal resolution was obtained at each spectral range. In addition, rapid differential acquisition with high-repetition rate operation effectively suppressed the intensity fluctuation of SC probe, allowing the detection of transient signals as small as $10^{-4}$. The developed platform provides single-µm scale spatial resolution and supports both transmission and reflection geometries covering both transparent and opaque samples, where the performance was demonstrated using multilayer and monolayer $MoS_2$. Importantly, the present approach is not limited to Ti:sapphire oscillators but can be extended to other ultrafast solid-state or fiber oscillators by using the tailored PCF for the different wavelength [21]. These results establish the proposed system as a versatile and accessible platform for investigating the local photoexcited dynamics of low-dimensional and nanoscale materials.

## Funding

Konica Minolta Science and Technology Encouragement Foundation; the Mitsubishi Foundation; the Murata Science Foundation; the Asahi Glass Foundation; the MEXT Leading Initiative for Excellent Young Researchers (No. JPMXS0320220123); JSPS Grant-in-Aid for Challenging Research (Pioneering) (No. 22K18268); JSPS Grants-in-Aid for Scientific Research (Nos. 26K08190(C), 24K01277(B), 23H01939(B), 20H02549(B)); JSPS Fostering Joint International Research (No. 24KK0257) and JSPS Grant-in-Aid for Transformative Research Areas (A) (No. 26A203). BC. acknowledges support from Spanish CM "Talento Program César Nombela" (project No. 2023-T1/TEC-28968), and the AEI for RYC2024-049194-I, PID2024-155345NA-I00, and CEX2024-001445-S)

## Acknowledgement

The authors gratefully acknowledge Prof. Yong Xie, Prof. Andrés Castellanos, Thomas Pucher, Qianjie Lei, and Jiahao Kang for their generous support in fabricating the CVD monolayer sample used in this study.

## Disclosures

The authors declare no conflicts of interest.

## Data Availability

Data underlying the results presented in this paper are not publicly available at this time but may be obtained from the authors upon reasonable request.